\documentclass[12pt,preprint]{aastex}

\slugcomment{Accepted, ApJ, March 12, 2010}
\shorttitle{Inclination Change of TrES-2b}
\shortauthors{Scuderi et al.}

\begin{document}

\title{On the Apparent Orbital Inclination Change of the Extrasolar Transiting Planet TrES-2b}

\author{Louis J. Scuderi, Jason A. Dittmann, Jared R. Males, Elizabeth M. Green, \& Laird M. Close}
\affil{Steward Observatory, University of Arizona, Tucson, AZ 85721}

\begin{abstract}
On June 15, 2009 UT the transit of TrES-2b was detected using the University of Arizona's 1.55 meter Kuiper Telescope with 2.0-2.5 millimag RMS accuracy in the I-band. We find a central transit time of $T_c = 2454997.76286 \pm0.00035$ HJD, an orbital period of $P =  2.4706127 \pm 0.0000009$ days, and an inclination angle of $i = 83^{\circ}.92 \pm 0.05$, which is consistent with our re-fit of the original I-band light curve of O'Donovan et al. (2006) where we find $i = 83^{\circ}.84 \pm0.05$. We calculate an insignificant inclination change of $\Delta i = -0^{\circ}.08 \pm 0.07$ over the last 3 years, and as such, our observations rule out, at the $\sim 11 \sigma$ level, the apparent change of orbital inclination to $i_{predicted} = 83^{\circ}.35 \pm0.1$ as predicted by Mislis \& Schmitt (2009) and Mislis et al. (2010) for our epoch. Moreover, our analysis of a recently published Kepler Space Telescope light curve (Gilliland et al. 2010) for TrES-2b finds an inclination of $i = 83^{\circ}.91 \pm0.03$ for a similar epoch. These Kepler results definitively rule out change in $i$ as a function of time. Indeed, we detect no significant changes in any of the orbital parameters of TrES-2b.
\end{abstract}

\section{Introduction}

Transiting extrasolar planetary systems are relatively rare with only 69 discovered extrasolar planets known to transit their host stars\footnote{http://exoplanet.eu}. These transits allow for direct measurement of certain properties of planets and their stellar hosts such as planetary radius, the inclination angle of the planetary system, limb darkening of the host star, and features on the surface of the host star. The exact shape of transit light curves depends heavily on all of these parameters (Mandel \& Agol 2002). Though this is a good method for detecting planets and it can provide an excellent characterization of the system, the sensitivity to very low radii (hence masses) is limited. While facilities such as the Kepler Space Telescope may soon be able to detect planets as small as or smaller than Earth (Borucki et al. 2008), there are currently only five known extrasolar transiting planets as small as or smaller than Neptune in mass\footnotemark[\value{footnote}]; GJ 436b with a mass of $\sim 21 M_\earth$ (Butler et al. 2004), HAT-P-11b with a mass of  $\sim 25 M_\earth$ (Bakos et al. 2009, Dittmann et al. 2009a), CoRoT-7b with a mass of $\sim 4.8 M_\earth$ (Queloz et al. 2009), GJ 1214b with a mass of $\sim6.55 M_\earth$ (Charbonneau et al. 2009), and the recently discovered Kepler-4b with a mass of $\sim24.5 M_\earth$ (Borucki et al. 2010).

This limitation is not nearly as absolute as it may seem, however. Mirlada-Escud\'{e} (2002) notes that even a relatively small secondary planet in a multiple planet system can induce significant observable changes in the orbital parameters of the transiting planet, most notably in the duration of the transit and in the inclination angle. Moreover, as stated in Ribas et al. (2008), grazing or near-grazing transits are especially sensitive to this type of perturbation. Several systems have been thoroughly examined for this type of perturbation. In the case of the elliptical orbit of GJ 436b there have been no positive signs of perturbations (Ribas et al. 2008, Alonso et al. 2008, Bean et al. 2008). The equally elliptical orbit of HAT-P-11b also shows no signs of clear variations (Dittmann et al. 2009a).

One important consideration in the search for changes in the orbital parameters of transiting extrasolar planets is the consistency of observations. Czesla et al. (2009) note that transit light curve morphology has a strong color dependence because of wavelength-dependent limb darkening laws (as observed on the Sun, for example). They further state that stellar activity may severely effect the morphology of the light curve. Rabus et al. (2009) and Dittmann et al. (2009b) show that this is indeed the case with the detection of starspot activity on the surface of transiting extrasolar planet host star TrES-1 during consecutive transits of TrES-1b.

TrES-2b is a $\sim1.2$ Jupiter-mass planet in a $\sim0.037$ AU radius orbit around the G0V star TrES-2,
which has approximately the same radius and mass as the Sun (Southworth 2009). TrES-2b was discovered via the transit method by O'Donovan et al. (2006) to have a radius of approximately 1.24 times that of Jupiter and an inclination angle of $i = 83^{\circ}.90 \pm 0.22$, which implies a near-grazing transit (with a grazing transit having a maximum inclination of $i = 83^{\circ}.417$ and a minimum of $i = 81^{\circ}.52$ as noted by Mislis \& Schmitt 2009).

These discovery observations were taken on the Sleuth telescope (Palomar Observatory, California), and the Planet Search Survey Telescope (Lowell Observatory, Arizona) in both $r$ and  $R$ filters. In order to obtain absolute photometry and colors of TrES-2, O'Donovan et al. (2006) observed the host star in Johnson $UBV$ and Cousins $R$ filters on the nights of 29 and 30 August 2006. A new set of rapid cadence, high-precision observations was taken in the $I$ filter ($\lambda \approx 806$ nm, $\Delta \lambda \approx 150$ nm) with the 0.82 meter IAC80 telescope at the Observatorio del Teide in Tenerife, Spain on 10 August 2006 UT. Simultaneous observations with lower cadence were carried out on this telescope in the $B$ filter and on the TELAST 0.35 meter telescope also at the Observatorio del Teide. Later observations by Holman et al. (2007) between epochs 13 and 34 from the ephemeris calculated by O'Donovan et al. (2006) at the 1.2 meter telescope of the Fred L. Whipple Observatory with an SDSS z-band filter  improved upon the parameters of the system. Raetz et al. (2009) further improved on the ephemeris of the system with a series of I-band observations of nine transits between epochs 87 and 318 at the 0.25 meter auxiliary telescope of the University Observatory Jena in Gro\ss schwabhausen, Germany.

Mislis \& Schmitt (2009) observed the system on 20 May and 18 September 2008 UT (epochs 263 and 312 respectively)  with the 1.2 meter Oskar L\"uhning telescope at Hamburg Observatory. The first set of observations were taken without a filter and the second was taken with an i-band filter ($\lambda \approx 775$ nm, $\Delta \lambda \approx 100$ nm). These observations suggested that the duration of the transit had decreased from the z-band Holman et al. (2007) observations by $\sim3.16$ minutes. This raised the possibility of another planet in the system perturbing the orbit of TrES-2b and causing a decrease in inclination (closer to a grazing transit). Furthermore, a follow up study by Mislis et al. (2010) confirmed their previous results and even suggested that TrES-2b is no longer fully transiting the parent star and is steadily decreasing its inclination. A follow-up observation to confirm such an inclination change was a major motivation for this paper.

\section{Observations \& Reductions}

We observed the transit of TrES-2b on 15 June, 2009 UT using the University of Arizona's 61 inch (1.55 meter) Kuiper telescope on Mt.\ Bigelow, Arizona with the Mont4k CCD, binned 3x3 to 0.43$^{\prime\prime}$/pixel. Our observations were taken only seven orbits after the last observations of Mislis et al. (2010) and the orbit immediately after the end of the Gilliland et al. (2010) data set. A serious concern in observing extrasolar planetary transits is the choice of filter to use. In this situation, when trying to detect subtle changes in duration and depth with respect to previous observations, it is important to choose a filter which matches the filter used in the previous observations. The original transit observations of O'Donovan et al. (2006) used an I-band filter. After careful consideration we chose to use the Arizona I-band filter already installed on the Mont4k filter wheel in order to match the original observations as closely as possible with available equipment. This filter is similar to a Kron-Cousins I-band filter, and differs only by a slightly bluer central wavelength from the exact specifications of the Kron-Cousins I. The filters that defined the system were made of combinations of colored glass which were used with an RCA 31034-A GaAs phototube.  The bluer bandpass of the Arizona-I ensures that the effective bandpass of the combined Mont4k filter and detector is a better match to the standard Kron-Cousins I when the filter bandpass is convolved with the CCD response curve.

TrES-2 is a G0V star at a distance of $220 \pm10$ pc with an apparent V magnitude of $V=11.41$ (O'Donovan et al. 2006).  This makes observing in the broad I-band slightly more difficult than in a narrower band, as the star is approximately 10 times brighter than the rest of the nearby stars in the surrounding field. Though one solution to this problem might be to de-focus the telescope for longer integration times, this option is rendered unfeasible since TrES-2 is in the highly crowded \textquotedblleft Kepler" field and there is a nearby field star at a radial distance of $\sim 10\arcsec$ which might contaminate the photometry of TrES-2 if the FWHM of the PSF is expanded. Moreover, keeping a consistent focus offset with this specific autoguider setup is prohibitively difficult. Additionally, all curves presented have slightly shallower transits than is correct due to the crowed nature of this Þeld. In particular, there is a faint ($\Delta I \sim 3.66$) star $1.089\arcsec$ from TrES-2 (Daemgen et al. 2009) which is blended into all the transit curves in this paper. The effect of this is that $\frac{R_p}{R_*}$ is underestimated by $\sim 2.1\% $(Daemgen et al. 2009). The aforementioned nearby Þeld star ($\Delta I \sim 5.1$) at a distance of $\sim 10\arcsec$ from TrES-2 would only e?ect the larger beam used in the Kepler lightcurve. The e?ect of this is still minor, leading to an underestimate of $\frac{R_p}{R_*}$  by $< 1\%$ in the Kelpler curve with respect to the other curves presented. Moreover, this would have no e?ect on the duration of the Kepler transit. Hence our results are independent of this e?ect which, if anything, brings the Kelpler curve into closer agreement with our curve.

The conditions during observation were nearly photometric and moonless. A low thin layer of smoke produced by a fire about 80 miles to the southwest near Kitt Peak was below the elevation of the telescope. There was a slight (with an average change of $\sim$0.3\% of the TrES-2 flux and a maximum of $\sim0.8$\% ) parabolic subtraction applied to the light curve to correct for the differential atmospheric extinction between the bluer G0V target and the redder reference stars used. After this simple correction our pre-ingress and post egress timeseries points were flat and constant in flux yielding a high quality light curve. We were able to use exposure times of 17 seconds, and by windowing the CCD to 1364x1090 pixels, we were able to decrease overhead time to just under 13.5 seconds for a total sampling period of approximately 30.5 seconds. Typical seeing ranged from 1$^{\prime\prime}$.1 to 1$^{\prime\prime}$.3 and there were less than four pixels ($< 1\arcsec.7$) of wander for any given star over the whole dataset due to autoguiding. The airmass of the target star ranged from $\sec{z} = 1.4$ at the beginning of observations to 1.1 at the end.

Each of the 535 images were bias-subtracted, flat-fielded, and bad pixel-cleaned in the usual manner. Before and after the TrES-2b light curve observations, six 600 second I-band fringe frames were taken on empty regions of sky. Between each exposure, the telescope was dithered between $15^{\prime\prime}$ and $20^{\prime\prime}$ west.  These were combined to produce a fringe frame for image correction. Upon further investigation, however, it was discovered that over the maximum pixel wander of each star there is only a maximum difference of $\lesssim10$ counts in the fringe pattern. This effect is negligible next to the typical 30-50,000 counts/peak pixel from TrES-2.

\begin{figure}[htp]
\centering
\includegraphics[scale=0.5]{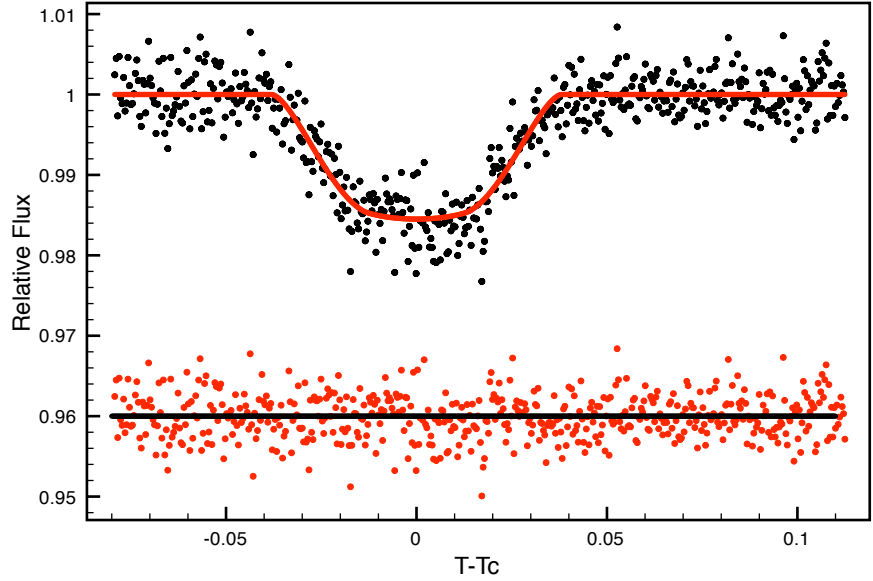}
\caption[Transit light curve with $\chi^2$ fit and residuals to fit]
{I-band time series photometry of the TrES-2b transit observed on June 15, 2009 UT. We show our best fit (reduced $\chi_\nu^2 = 1.2$) as a solid red curve. The 2.5 mmag rms residuals of the fit are shown below (red points). No data points were removed from this light curve or fit. Table 1 gives a summary of our fit parameters.}
\label{lightcurve}
\end{figure}

\begin{figure}[htp]
\centering
\includegraphics[scale=0.5]{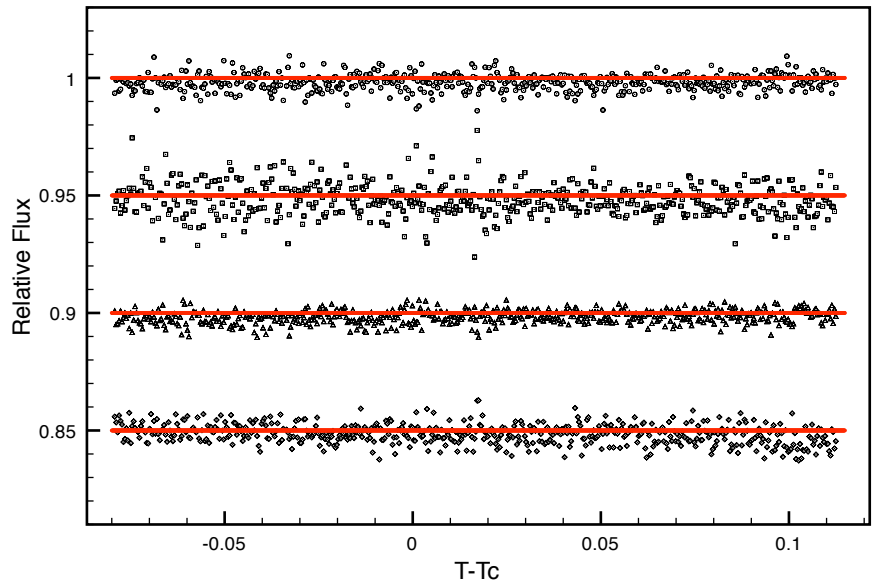}
\caption[Calibrator star light curves]
{Standard star light curves each normalized by the sum of the remaining three calibrator stars. No data points were removed from any of these light curves. Each curve is offset for clarity.}
\label{reference}
\end{figure}

Aperture photometry was performed using the aperture photometry task PHOT in the IRAF DAOPHOT package.\footnote{IRAF is distributed by the National Optical Astronomy Observatories, which are operated by the Association of Universities for Research in Astronomy, Inc., under cooperative agreement with the National Science Foundation.} A 4$\arcsec$.3 aperture radius (corresponding to 10.0 pixels-see light curve shown in figure \ref{lightcurve}) was adopted because it produced the lowest scatter in the resultant lightcurve and smallest error terms for each data point, though apertures of radius 3$\arcsec$.44 (8 pixels) and 5$\arcsec$.16 (12 pixels) were also analyzed and yielded nearly identical results. Several combinations of reference stars were considered, but four were selected for the final reduction because of the lack of linear trends and low RMS scatter.  The four reference stars used were at distances of $67^{\prime\prime}, 149^{\prime\prime}, 157^{\prime\prime},$ and $194^{\prime\prime}$ from TrES-2. The time series light curves are shown for each of these reference stars in figure \ref{reference}.

\begin{figure}[htp]
\centering
\includegraphics[scale=0.5]{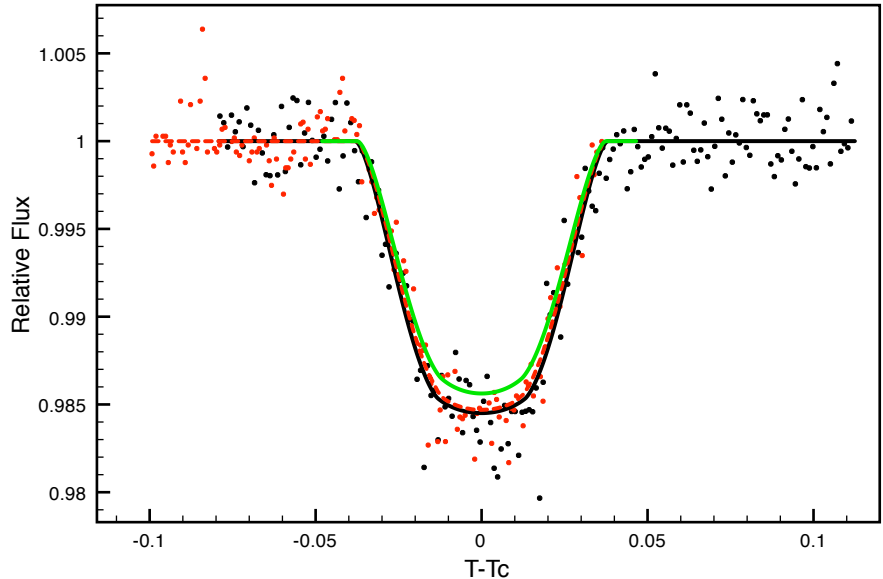}
\caption[Data from O'Donovan et al. (2006) and this paper with fits to both.]
{Our I-band data from the June 15, 2009 UT transit observed in this paper binned by three data points (black points). We also plot the I-band transit of  O'Donovan et al. (2006) (red points) with new minimized $\chi^2$ fits shown to both data sets (fit to our data: solid black line, our fit to the O'Donovan et al. (2006) data: dashed red line, closely overlapping the black line). These two fit curves are remarkably similar even though the two transits were separated by almost three years (421 orbital periods). We also plot the results of our fit to the very high S/N Kepler model presented by Gilliland et al. (2010), (green line). Though this Kepler model shows a slight inconsistency in transit depth it is still within our errorbars and remains much deeper and longer than the predictions of Mislis \& Schmitt (2009) or Mislis et al. (2010). We also note that the duration of all three transits are nearly identical, hence all three curves have nearly identical inclinations.}
\label{overplot}
\end{figure}

We applied no sigma clipping rejection to the reference stars or TrES-2b; all datapoints were used in the analysis. The final light curve for TrES-2 was normalized by division of the weighted average of the four reference stars. The unbinned residual time series in figure \ref{lightcurve} has a photometric RMS range of 2.0 to 2.5 mmag.  This is just slightly larger than typical errors for the Mont4k on the 61-inch (Kuiper) telescope for high S/N images (Dittmann et al. 2009a, Dittmann et al. 2009b). When applying a static binning over every 3 data points, our RMS scatter becomes similar to the I-band transit of O'Donovan et al. (2006) as seen in figure \ref{overplot}.

\section{Analysis}

\subsection{Fitting the I Band Data}
All of the transit light curves produced were fit using the $\chi ^{2}$ minimization method prescribed by Mandel and Agol (2002). The quadratic limb darkening coefficients for the I-band filter were adopted from Claret (2000) as 0.2119 and 0.3434 respectively. To arrive at these values, we adopted stellar parameters close to those in O'Donnovan et al. (2006) which matched with the discrete stellar parameters in Claret (2000) of $T_{eff} = 6000$ K, $log{g} = 4.5$, and a turbulent velocity of $V_T = 2.0 km s^{-1}$. In order to detect any transit duration variations as suggested by Mislis \& Schmitt (2009) we allowed the center of transit, $T_c$, and the inclination angle of the orbit, $i$, to vary. The parameters used in our fit were adopted from O'Donovan et al. (2006) in order to be consistent with those used in the analyses of Holman et al. (2007) and Mislis \& Schmitt  (2009) These parameters are shown in Table 1. We note that the purpose of this section of the paper is not to re-derive all of the parameters of the transit, but to understand if our transit is consistent with the predictions of Mislis \& Schmitt (2009) or O'Donovan et al. (2006).

We minimized the reduced $\chi_\nu ^{2}$ to 1.2 (see figure \ref{lightcurve} and Table 2) fitting these two parameters to find a center of transit time $T_c = 2454997.76286 \pm0.00035$ HJD and an inclination angle $i = 83^{\circ}.92\pm 0.05$ (the flux uncertainty for each datapoint in the $\chi_\nu ^{2}$ fit was calculated from the propagation of photometric errors determined with the PHOT task in IRAF). Allowing all parameters to vary results in a negligible change in depth and a reduced $\chi_\nu ^{2}$ of 1.17, which is an insignificant improvement. From these parameters, we were able to calculate a total transit duration of $T_D = 109.6 \pm0.5$ minutes from first to fourth contact. While this is consistent at the $\sim 2\sigma$ level with the duration presented by Holman et al. (2007) ($T_D = 110.4 \pm 1.2$ minutes) , it is inconsistent with the measurement of Mislis \& Schmitt (2009) ($T_D \approx 107.2$ minutes) at the $\sim 6\sigma$ level, and with the Mislis et al. (2010) measurement ($T_D \approx 104.52$ minutes) at the $\sim 10\sigma$ level. Each of our $1\sigma$ uncertainties were estimated by Monte-Carlo simulations of 1000 simulated datasets with the same RMS scatter as the original data (as described in Dittmann et al. 2009 a,b).

We also re-fit the original I-band transit observation data taken by O'Donovan et al. (2006) and find that this fit is consistent with our best fit to our own I-band data (figure \ref{overplot}). We minimized $\chi^2$ and let the center of transit time, $T_c$, and inclination angle, $i$ vary, and we adopted all of the same parameters used in the fit to our own data (see table 1). We find an inclination angle $i = 83^{\circ}.84 \pm 0.05$.

\subsection{New results from the Kepler mission}
\begin{figure}[htp]
\centering
\includegraphics[scale=0.5]{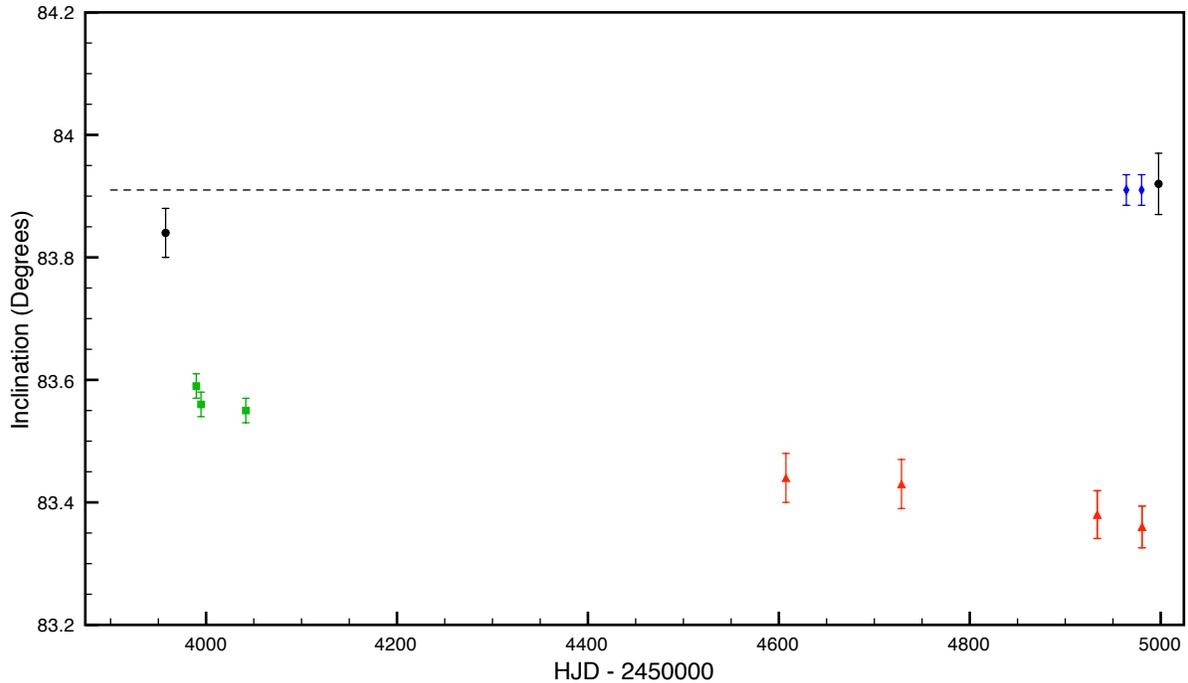}
\caption[Inclination angle for TrES-2 for various epochs.]
{The measured inclination angle for transit of TrES-2 over all observations discussed in this paper. The Kepler data, from Gilliland et al. (2010) is represented by the blue diamonds (error bars are the size of the points) with two points indicating the beginning and end of the Kepler observing period, the data presented by Holman et al. (2007) is represented as green boxes, and the data presented in both Mislis \& Schmitt (2009) and Mislis et al. (2010) is shown in red triangles. The data presented in this paper and our re-fit of the O'Donovan et al. (2006) discovery data are presented as black circles. The dashed line represents the inclination observed in the Kepler data. We note that both our observations and the O'Donovan et al. (2006) observations agree well with the high S/N Kepler data and suggest no change in inclination.}
\label{inclinations}
\end{figure}

\begin{figure}[htp]
\centering
\includegraphics[scale=0.5]{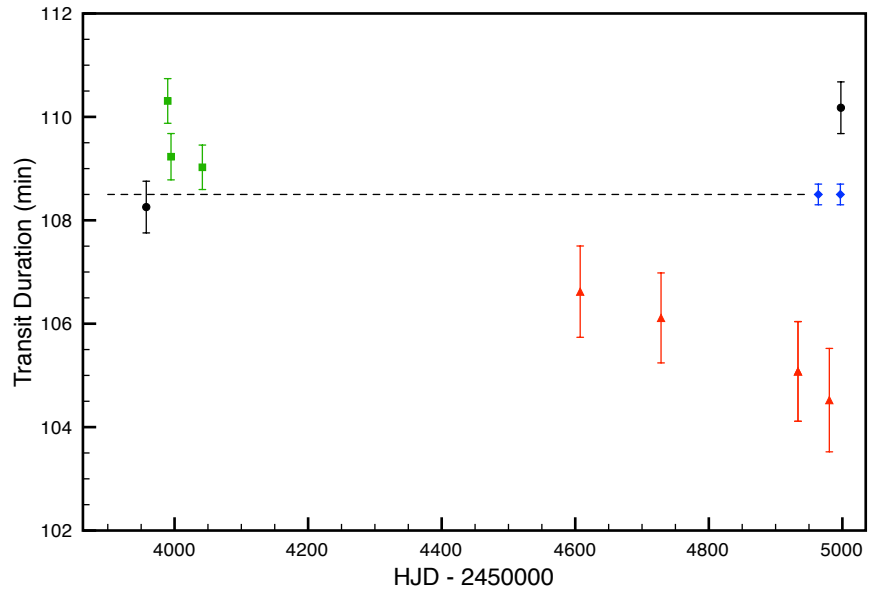}
\caption[Transit Duration Times for TrES-2 for various epochs.]
{The measured transit duration time for transits of TrES-2 over several sets of observations. The data points are indexed as in figure \ref{inclinations}. We note that the transit duration from O'Donovan et al. (2006), Holman et al. (2007), and Gilliland et al. (2010) are consistent with our measurements at the $\lesssim 2 \sigma$ level. Again, we cannot reconcile the Mislis data with the higher S/N Kepler or 1.55 meter data.}
\label{durations}
\end{figure}

Gilliland et al. (2010) recently released Kepler Space Telescope data of multiple transits of TrES-2b during the first quarter of observations, which ended on 15 June 2009 UT and spanned the time period HJD 2454964.00 to HJD 2454997.49. We note that this time period exactly precedes the transit presented in this paper and includes one of the transits presented in Mislis et al. (2010). Though Gilliland et al. (2010) do not report the results of their multiple phase-folded transit fit, instead only using the data as a demonstration of principle, we were able to extract the fit curve ourselves (see figure \ref{overplot}). Since the Kepler spacecraft has a wide spectral response, spanning both the V and R filters (Koch et al. 2009), we attempted a fit to their model with several limb darkening (LD) parameters adapted from Claret (2000). With R-band LD coefficients of 0.3374 and 0.3238, the inclination was found to be $i = 83^{\circ}.890 \pm0.001$. With V-band LD coefficients of 0.4369 and 0.2952, the inclination was found to be $i = 83^{\circ}.940 \pm0.001$. We also took the average LD coefficients between the two bands and found an inclination angle of $i = 83^{\circ}.91 \pm0.03$. We note that these values are consistent with experimental error with both our $i =83^{\circ}.92 \pm0.05$ and that of O'Donovan et al. (2006) (see figures \ref{inclinations} and \ref{durations}).

This analysis shows no change within experimental errors between transits taken nearly 3 years apart. We are not completely sure why Mislis \& Schmitt (2009) and Mislis et al. (2010) systematically find lower inclinations, but this might be due to poor seeing, higher airmass ranges, and the use of a single reference star in Mislis et al. (2010). In any case, we find the inclination is not changing using high quality ground and space-based photometry.

\subsection{Transit Duration Measurements}
Because TrES-2b is a grazing transit, a slight change in inclination angle can significantly change the transit duration. It is also easier to determine the duration of the transit from the curve than the inclination angle of the planet because the duration depends on when the planet leaves the disc of the star and is clearly visible on the light curve while the inclination is determined from the shape of the ingress and egress portions of the curve and is more significantly affected by photometric errors. We followed the same procedure above, and measured the transit duration for each data set. These results are shown in figure \ref{durations}. We find that our Kepler fit and 1.55 meter duration measurements are consistent with that of O'Donovan et al. (2006) and Holman et al. (2007). We suggest that the change found by Mislis \& Schmitt (2009) and Mislis et al. (2010) is due to systematic effects and not to any real physical change in transit duration.

\subsection{What if the orbital inclination angle is changing?}
\begin{figure}[htp]
\centering
\includegraphics[scale=0.5]{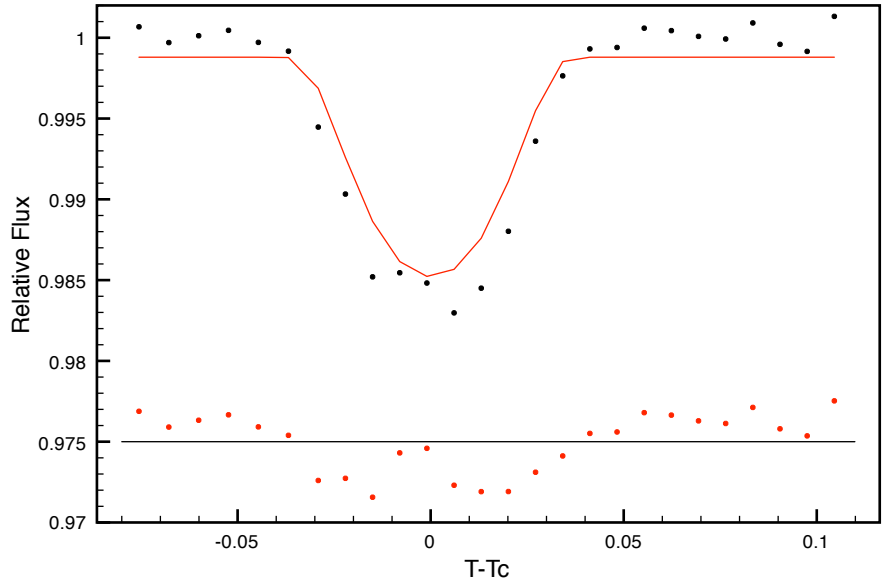}
\caption[Light curve with forced fit adopted from Mislis \& Schmitt (2009).]
{Our transit observations binned by 20 data points from the transit of 15 June, 2009 UT with a fit to a forced inclination angle of $i = 83^{\circ}.35$ as predicted for our epoch by Mislis \& Schmitt (2009). The fit is very poor with a large reduced $\chi_\nu ^2 = 6.9$, and failed to normalize out of transit points to unity or to match the full depth and shape of the observed light curve. Below are the residuals to the fit. It is clear that we can reject the possibility that $i = 83^{\circ}.35$ during our observation.}
\label{forcedfit}
\end{figure}

As a final test to determine whether the orbital inclination angle is changing, we will take the trend suggested by Mislis \& Schmitt (2009) and use it to predict the transit curve for our observation. Their trend suggests that at the time of our observation, TrES-2b had an inclination angle of $i = 83^{\circ}.35$. Fixing the model at this angle, and fitting for the transit time only, we find a reduced $\chi^{2}$ of 6.9. This is definitely a poor fit to the data and it is clear that the residuals have a very obvious trend lingering from the transit (See figure \ref{forcedfit}). Also, it is impossible to fit $i = 83^{\circ}.35$ to the Gilliland et al. (2010) Kepler model due to the extremely high signal-to-noise of that data set. Only a much higher value of $i = 83^{\circ}.91 \pm 0.03$ can fit the Kepler data set.

\begin{figure}[htp]
\centering
\includegraphics[scale=0.5]{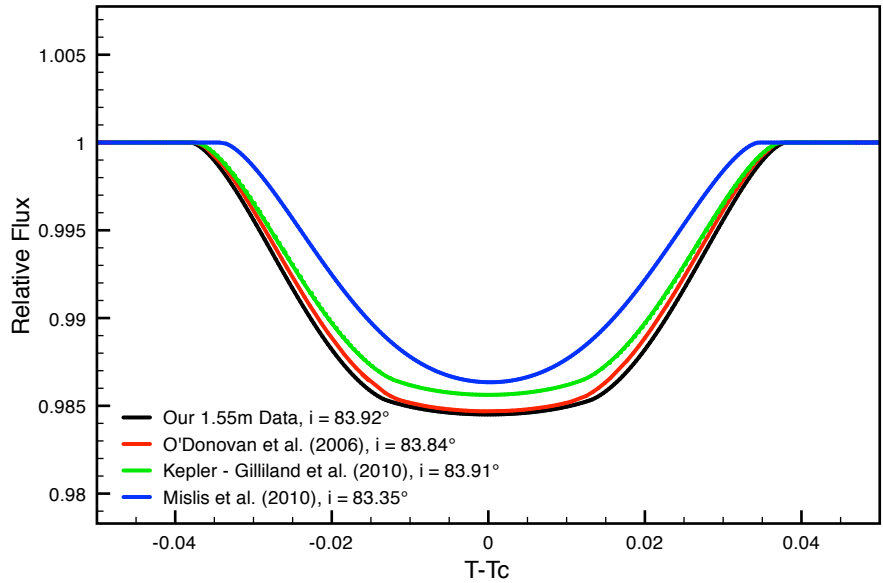}
\caption[Comparison of various model fits from O'Donovan et al. (2006), Mislis et al. (2010), Gilliland et al. (2010), and this work.]
{Comparison of various model fits from O'Donovan et al. (2006), Gilliland et al. (2010), this work, and Mislis et al. (2010). While the first three models seem to agree to within error tolerances, the Mislis et al. (2010) model diverges greatly on transit duration, ingress and egress shape, and depth. Though the Gilliland et al. (2010) model shows a slightly shallower transit, it agrees extremely well with our data and our re-fit of the O'Donovan et al. (2006) data in duration and ingress and egress shape, and hence inclination. This suggests no change in orbital parameters since the discovery by O'Donovan et al. (2006).}
\label{models}
\end{figure}

The orbital inclination of $i = 83^{\circ}.35$ suggested by Mislis \& Schmitt  (2009) also means that not only would the transit be shorter, but it would also be shallower. They predicted that in October 2008, TrES-2b would no longer be fully transiting, ie - a portion of the disc would miss occulting the host star entirely. This means that the transit should not be as deep as it was when the planet was discovered. In fact, Mislis et al. (2010) claim to have seen and measured this effect in the spring of 2009. However, our data set was taken after the Mislis et al. (2010) data set, so if the proposed inclination change is real and not due to systematic noise, we should be able to continue to see this trend in our data set. To investigate this, we overlay our I-band data with the I-band data of O'Donovan et al. (2006). If TrES-2b really has passed the first inclination threshold, this should be apparent in our data. As can be seen in figure \ref{models}, we see no evidence for a change in transit depth over the past 3 years of observations. In fact, when comparing the model fits from O'Donnovan et al. (2006), Mislis et al (2010), and both the fit to the data presented in this paper as well as our fit to the data presented in Gilliland et al. (2010), we find in figure \ref{models}  that the Mislis et al. (2010) model is extremely inconsistent with the other three models.

\section{Discussion}
The main goal of this paper is to compare our measured orbital parameters to those of Mislis \& Schmitt (2009) and O'Donovan et al. (2006) to see if the orbital parameters of the system are, in fact, changing as suggested. Projecting the ephemeris of $T_c(HJD) = (2453957.63492 + E á 2.470614)$ days (Raetz et al. 2009) forward to the current epoch (421 orbits later), we find that the center of transit $T_c$ from our best fit curve is $48 \pm 38$ seconds early, which is entirely consistent with the ephemeris of Raetz et al. (2009). Therefore we agree with Mislis et al. (2010) that there are no transit timing variations for TrES-2b.

Because TrES-2b is a grazing transit, we were able to investigate changes in transit duration that would be a result of a changing inclination angle. Combined, we have found that the transit duration, $T_D = 109.6 \pm 1.5$ minutes and the inclination angle $i =  83.92^{\circ} \pm 0.05$ are completely consistent with the parameters published in O'Donovan et al. (2006) ($i = 83^{\circ}.90 \pm 0.22$). However, our own reanalysis of the 2006 I-band light curve shown in figure \ref{overplot} suggests an inclination of $i = 83.84 \pm 0.05$. Therefore, the observed change of inclination from our re-fit of the O'Donovan et al. (2006) I-band data is an insignificant $\Delta i = -0^{\circ}.080 \pm0.071$ over the last $\sim 3$ years, inconsistent with that suggested by Mislis \& Schmitt (2009) and Mislis et al. (2010).

Kepler Space Telescope data presented by Gilliland et al. (2010) also casts doubt on the accuracy of claims made by Mislis \& Schmitt (2009) and Mislis et al. (2010). When fitting the Gililand et al. (2010) model via our own analysis process in a manner consistent with previous analyses, we find a range of inclination from $i = 83^{\circ}.890 \pm0.001$ (using R-band limb darkening coefficients) to $i = 83^{\circ}.940 \pm0.001$ (using V-band LD coefficients). These observations were taken concurrently with those presented in Mislis et al. (2010), and seem to rule out any inclination change from that presented in the discovery paper by O'Donovan et al. (2006). Moreover, we find no significant change in duration from the Kepler data (with a duration of $\approx 108.5 \pm 0.2$ minutes).

\section{Conclusions}

We investigated if our I-band transit light curve is significantly different from an I-band curve taken $\sim3$ years ago by O'Donovan et al. (2006). We report that the inclination and transit duration of the transiting extrasolar planet TrES-2b has not changed significantly ($\Delta i = -0^{\circ}.080 \pm 0.071$) from the original values given in O'Donovan et al. (2006). We find a central transit time of $T_c = 2454997.76286 \pm0.00035$ HJD from a best fit to our data. We estimate that our I band transit had a duration of $109.6 \pm0.5$ minutes with an inclination angle of $i = 83.92^{\circ}\pm0.05$ and a period of $P = 2.4706127 \pm 0.0000009$ days. Our new, slightly more accurate ephemeris calculated from these parameters is $T_c (E) = 2453957.6349133 + E \times 2.4706127$. Moreover, when fitting to a model of data obtained by the Kepler Space Telescope and presented in Gilliland et al. (2010), we find no significant change in system parameters, nor any evidence of a trend in inclination. As a whole, these findings are entirely consistent with the ephemeris calculated by Raetz et al. (2009), and rule out the possibility of any large duration change or orbital parameter variation as suggested by Mislis \& Schmitt (2009) or Mislis et al. (2010).

\acknowledgments

We would like to thank the anonymous referee whose comments have led to a much better paper. We would like to thank the Arizona NASA Space Grant program for funding this work. LMC is supported by a NSF Career award and the NASA Origins program.

{\it Facilities:} \facility{Kuiper 1.55m}.

\clearpage
{\bf References}\\\

Alonso, R., Barbieri, M., Rabus, M., et al., 2008, A\&A, 487, L5 \\

Bakos, G. \'A., Torres, G. P\'al, A., et al., 2009, ApJ submitted (arXiv:0901.0282v1) \\

Bean, J. L., Benedict,  G. F., Charbonneau, D., et al., A\&A, 486, 1039 \\

Borucki, W., Koch, D., Basri, G., et al., 2008, Exoplanets: Detection, Formation and Dynamics, Proceedings of the International Astronomical Union, IAU Symposium, Vol. 249, 17\\

Borucki, W., Koch, D., Brown, T. et al., 2010, ApJL submitted (arXiv: 1001.0604v1)\\

Butler, P., Vogt, S., Marcy, G., et al., 2004, ApJ, 617 L580 \\

Charbonneau, D., Berta, Z., Irwin, J., et al. 2009, Nature 462, 891-894\\

Claret, A. 2000, A\&A, 428, 1001\\

Czesla, S., Huber, K.F., Wolter, U., et al. 2009, ApJ accepted\\

Daemgen, S., Hormuth, F., Brandner, W., et al. 2009, A\&A, 498, 567\\

Dittmann, J.A., Close, L.M., Green, E.M., Scuderi, L.J., Males, J.R. 2009a, ApJ, 699, L48.\\

Dittmann, J. A., Close, L.M., Green, E.M., Fenwick, M. 2009b, ApJ, 701, 756-763\\

Gilliland, R.L., Jenkins, J.M., Borucki, W.J. et al., 2009, ApJL in press (arXiv: 1001.0142v1)\\

Holman, M.J., Winn, J.N., Latham, D.W., et al., 2007, ApJ, 664, 1185\\

Koch, D., Borucki, W., Basri, G., et al., 2009 ApJL, in press, (arXiv: 1001.0268v1) \\

Mandel, K., \& Agol, E. 2002, ApJ, 580, L171\\

Miralda-Escud\'e, J., 2002, ApJ, 564, 1019\\

Mislis, D., Schmitt, J.H.M.M., 2009, A\&A accepted (arXiv:0905.4030v1)\\

Mislis, D., Schr\"oter, S., Schmitt, J.H.M.M., Cordes, O., Rief, K., 2010, A\&A, submitted (arXiv:0912.4428v2)\\

O'Donovan, F. T., Charbonneau, D., Mandushev, G., et al. 2006, ApJ, 651, L61 \\

Queloz, D., Bouchy, F., Moutiou, C., et al. 2009, A\&A, 506, 303\\

Rabus, M., Alonso, R., Belmonte, J.A. et al., 2009, A\&A, in press (arXiv:0812.1799v1) \\

Raetz, St., Mugrauer, M., Schmidt, T.O.B., et al., 2009, Astronomische Nachrichten, 330, 459 \\

Ribas, I., Font-Ribera, A., Beaulieu, J.P. 2008, ApJ, 677 L59\\

Southworth, J., 2009, MNRAS, 394, 272 \\

Winn, J. N. 2009, in IAU Symposium, Vol. 253, IAU Symposium, 99\\

\clearpage

\begin{table}
\begin{center}
\caption{Parameters of the TrES-2 system\label{tbl-1}}
\begin{tabular}{crrrrr}
\tableline\tableline
  Parameter [units] & Epoch & Filter & Value & Reference \\
  \tableline
   $R_{p}$ [$R_{J}$] & N/A & N/A &$1.24 \pm 0.06$ & b\\ 
   $R_*$ [$R_\sun$] & N/A & N/A & $1.00\pm0.06$ & b\\ 
   $R_{p}/R_{*}$ & N/A & N/A & 0.125 $\pm$ 0.028 & b\\ 
   $M_{p}$ [$M_{Jup}$] & N/A & N/A & $1.280\pm 0.069$ &b\\
   $T_c (E)$ [HJD] & 0 & $I,B,r,R$ & $2453957.6358 + E \times 2.47063$ & b\\
   $T_c (E)$ [HJD] & 34 & SDSS $z$ & $2453957.63479 + E \times 2.470621$ & c\\ 
   $T_c (E)$ [HJD] & 312 & No Filter, $i$ & $2453975.63403 + E \times 2.4706265$ & d\\ 
   $T_c (E)$ [HJD] & 318 & $I$ & $2453957.63492 + E \times 2.470614$ & e\\
   $P$ [days] & 0 & $I,B,r,R$ & 2.47063 $\pm$ 0.00001 & b\\
   $P$ [days] & 34 & SDSS $z$ & 2.470621 $\pm$ 0.000017 & c \\
   $P$ [days] & 312 & No Filter, $i$ & 2.4706265 $\pm$ 0.00001 & d \\
   $P$ [days] & 318 & $I$ & 2.470614 $\pm$ 0.00000100 & e\\
   $i$ [deg] & 0 & $I,B,r,R$ & $83.90 \pm 0.22$ & b\\
   $i$ [deg] & 0 & $I$ & $83.84 \pm 0.05$ & f\\
   $i$ [deg] & 34 & SDSS $z$ & $83.57 \pm 0.14$ & d\\
   $i$ [deg] & 312 & No Filter, $i$ & $83.430 \pm 0.036$ & e\\
   $i$ [deg] & 395,414 & $v,b,y,I$ & $83.36 \pm 0.03$ & g\\
   $i$ [deg] & 407-420 & Kepler & $83.91 \pm 0.03$ & h\\
  \tableline
  $T_c$ (421) [HJD] & 421 & $I$ & 2454997.76286 $\pm$ 0.00035 & This work \\ 
  $P$ [days]$^*$ & 421 & $I$ & 2.4706127 $\pm$ 0.0000009 & This work \\
  $T_{c} (E)$ [HJD] & 421 & $I$ & $2453957.6349133 + E \times 2.4706127$ & This work \\ 
  $i$ [deg] & 421 & $I$ & 83.92 $\pm$ 0.05 & This work \\
  $\Delta i$ [deg] & 0-421 & $I-I$ & $-0.080 \pm 0.071$ & This work \\
  \tableline
\end{tabular}

\tablenotetext{*}{the new period value was calculated by a sigma weighted least square of all five $T_c$ values}
\tablenotetext{a}{Southworth (2009)}
\tablenotetext{b}{O'Donovan et al. (2006)}
\tablenotetext{c}{Holman et al. (2007)}
\tablenotetext{d}{Mislis \& Schmitt (2009)}
\tablenotetext{e}{Raetz et al. (2009)}
\tablenotetext{f}{This work, re-processed from O'Donovan et al. (2006) I-band data.}
\tablenotetext{g}{Mislis et al. (2010)}
\tablenotetext{h}{This work, analyzed from Gilliland et al. (2010) Kepler data.}

\end{center}
\end{table}

\clearpage

\clearpage

\begin{table}
\begin{center}
\caption{Sample data table of the June 15 2009 light curve presented in this work. The entire table is available online.\label{tbl-2}}
\begin{tabular}{crrr}
\tableline\tableline
  HJD - 2454990.0 & Relative Flux & Error in Flux \\
  \tableline
  7.683780 & 1.002221 & 0.004130 \\
  7.684168 & 1.004263 & 0.004125 \\
  7.684555 & 0.997113 & 0.004124 \\
  7.684945 & 1.004501 & 0.004130 \\
  7.685331 & 0.999326 & 0.004128 \\
  7.685718 & 0.998627 & 0.004125 \\
  7.686107 & 1.001890 & 0.004124 \\
  7.686489 & 0.997712 & 0.005230 \\
  7.686875 & 0.998710 & 0.004126 \\
  7.687262 & 1.004375 & 0.004125 \\
  7.687653 & 1.000874 & 0.004129 \\
  7.688040 & 0.998508 & 0.004125 \\
  7.688426 & 0.998659 & 0.005232 \\
  7.688812 & 0.999857 & 0.004130 \\
  ... & ... & ... \\
  \tableline
\end{tabular}

\end{center}
\end{table}

\clearpage

\clearpage

\end{document}